\documentstyle[twoside,fleqn,espcrc2,graphicx,isolatin1]{article} 

\title{SSOR Preconditioning of Improved Actions
  \thanks{Talk presented by N. Eicker.}}

\author{N.~Eicker\address{HLRZ, c/o Research Center Jülich, D-52425
    Jülich, Germany}, W.~Bietenholz$^{\rm a}$,
  A.~Frommer\address{Department of Mathematics, University of
    Wuppertal, D-42097 Wuppertal, Germany},
  H.~Hoeber\address{Department of Physics, University of Wuppertal,
    D-42097 Wuppertal, Germany}, Th.~Lippert$^{\rm a}$,
  K.~Schilling$^{\rm a,c}$}

\begin{document}

\begin{abstract}
  We generalize {\it l}\/ocal {\it l}\/exicographic SSOR
  preconditioning for the Sheikholeslami-Wohlert improved Wilson
  fermion action and the truncated perfect free fermion action.  In
  our test implementation we achieve performance gains as known from
  SSOR preconditioning of the standard Wilson fermion action.
\end{abstract}

\maketitle

\section{INTRODUCTION}

The standard Wilson fermion action of lattice QCD leads to
discretization-errors of $O(a)$ in lattice spacing, requiring
prohibitively fine lattice resolutions in the approach to the chiral
and continuum limits \cite{LUESCHER9605038}.  The present trend to
tackle this problem goes in two directions: ({\em i}\/) One approach
is based on {\em Symanzik's on-shell improvement} program, where
irrelevant $O(a)$ counter terms are added to both, lattice action
(Sheikholeslami-Wohlert-Wilson action SWA) and composite operators
\cite{CLOVER}.  The hope is to reach the continuum limit for a
specific observable ${\cal O}(a) = {\cal O}_{cont.} + c_2 a^2 + \dots
$ without $O(a)$ contamination.  ({\em ii}\/) Another promising ansatz
is based on so-called {\em perfect actions} that are located on
renormalized trajectories intersecting the critical surface in a fixed
point of a renormalization group transformation \cite{HASENFRATZ}.
Perfect actions are in principle free of cut-off effects. However,
they can only be realized approximatively as truncated perfect actions
(TPA).

Simulations of dynamical fermions within these schemes meet the
problem of the compute intensive solutions of the fermionic linear
system $Mx=\phi$, well known from traditional actions.  In the last
three years, a considerable acceleration of the inversion of the
standard Wilson fermion matrix has been achieved by introduction of
the BiCGStab algorithm \cite{FROMMER94} and novel parallel {\it
  l}\/ocal-{\it l}\/exicographic SSOR preconditioning techniques
\cite{FROMMER96}.  Obviously, the efforts should be combined, i.e.\ 
{\it ll}\/-SSOR generalized for SWAs and TPAs in order to gain their
full pay-off\footnote{In Ref.~\cite{JANSEN}, odd-even preconditioning
  has been applied to the Sheikholeslami-Wohlert-Wilson action.}.

In general, both SWA and TPA can be written in the form
\begin{equation}
M=A+B+C+\cdots,
\label{eq:shape}
\end{equation}
where $A$ represents diagonal blocks (containing $12\times 12$
sub-blocks), $B$ is a nearest-neighbor hopping term, $C$ contains
next-to-nearest-neighbor couplings. Usually next-next-nearest-neighbor
couplings are truncated.

Our key observation is that one can include into the {\it ll}\/-SSOR
process ({\em i}\/) the internal degrees of freedom of the block
diagonal term $A$ as arising in SWA and ({\em ii}\/)
next-to-nearest-neighbor terms $C$ as present in TPA.

\section{PRECONDITIONING SWA}
Preconditioning amounts to the replacement of $M$, $x$ and $\phi$ by
preconditioned quantities $\tilde{M}$, $\tilde{x}$ and $\tilde{\phi}$.
The aim is to transform the matrix such that the spectrum becomes
narrower, increasing the efficiency of the inversion.  The
matrix-vector multiplication is replaced by
\begin{equation}
  v_i=Mp_i \quad\Rightarrow
  \left\{
    \begin{array}{l}
      \mbox{solve}\quad Pz_i = p_i \\
      v_i=Mz_i
    \end{array}
  \right. .
  \label{eq:matvec_repl_1}
\end{equation}
$P$ represents the preconditioning matrix.  It can be decomposed into
a product of three regular matrices $P=RST$.  This allows to apply the
'Eisenstat-trick' \cite{EISENSTAT} using the identity $M=R+T-K$, with
a fourth regular matrix $K$.

Let us recall the essentials of {\it ll}\/-SSOR \cite{FROMMER96}: The
matrix $M$ is decomposed into a (block-) diagonal part $D$, and two
strictly upper and lower triangular parts $U$ and $L$, $ M=D-L-U$.
This can be achieved by {\it l}\/ocal {\it l}\/exicographic ordering,
described in \cite{FROMMER96}. The choices for $R$, $S$ and $T$ are
$R=\frac{1}{\omega} D - L$, $S=\left(\frac{2-\omega}{\omega} D
\right)^{-1}$ and $T=\frac{1}{\omega} D - U$. Here $\omega$ is an
over-relaxation parameter to be chosen appropriately.

These special choices of $R$, $S$ and $T$ simplify the task of solving
the linear equation in (\ref{eq:matvec_repl_1}). They lead to the
following replacement of the matrix-vector multiplication:
\begin{equation}
  \label{eq:matvec_repl_2}
  v_i=M p_i \quad\Rightarrow \left\{
  \begin{array}{l}
    \mbox{multiply by }D\\
    \mbox{solve backward}\\
    \mbox{solve forward}
  \end{array} \right . .
\end{equation}

For standard Wilson fermions, the block-diagonal term is given by $A
\propto \mathbf{1}$, which implies the natural choice of $D \propto
\mathbf{1}$ in the SSOR scheme.  Therefore, in
(\ref{eq:matvec_repl_2}) the multiplication by $D$ and the
multiplication by $D^{-1}$ in the forward-/backward-solve is readily
carried out.  In the multiplication with a diagonal block, the 12
color-spin elements in the vector $x$ are decoupled and can be treated
simultaneously.

The situation changes if $A$ is not a strict diagonal.  We have the
freedom to choose the splitting of $A$ into the diagonal term $D$ and
the upper and lower terms $U$ and $L$.  As efficient implementations
require to store $D^{-1}$, we thus can control the memory overhead.
However, depending on the choice of $D$ the elements of $x$ are
intermixed in the multiplication with a diagonal block.  Therefore
they can only be treated simultaneously, if we choose the diagonal
part as $D = A$, the choice with the largest memory overhead.  For any
other choice, SSOR subprocesses on the diagonal blocks have to be
introduced.

As a test we have implemented the {\it ll}\/-SSOR preconditioning
scheme within BiCGStab for SWA.  The diagonal part of the related
quark matrix contains four complex $3\times 3$ matrices $F_i$:
\begin{equation}
  \left(
    \begin{array}{cccc}
      {\mathbf 1}+F_1&F_2&F_3&F_4\\
      F_2^\dagger&{\mathbf 1}-F_1&F_4^\dagger&-F_3\\
      F_3&F_4&{\mathbf 1}+F_1&F_2\\
      F_4^\dagger&-F_3&F_2^\dagger&{\mathbf 1}-F_1\\
    \end{array}
  \right)
  \label{eq:structure}.
\end{equation}
This structure reduces the storage requirements by a factor of $4$ and
is well suited to a QCD optimized machine.

We tested the inverter in a $16^4$ pure gauge background at
$\beta=6.0$ for two choices of $D$, ({\em i}\/) the true diagonal
({\em true}\/) with twelve $1\times 1$ blocks and ({\em ii}\/) the
${\mathbf 1}\pm F_1$ blocks ({\em block}\/) as shown in
(\ref{eq:structure}).  The $c_{SW}$-parameter was chosen as $1.0$ and
$1.6$. We tested the algorithm for different values of $\kappa$ on $4$
field-configurations.  The tests were done on the 32-node
APE100/Quadrics Q4 in Wuppertal. We compare to unpreconditioned
BiCGStab rescaled by a factor $2$ to mimic odd-even preconditioning as
a reference.

\begin{figure}[htb]
  \includegraphics[angle=270,width=\columnwidth]{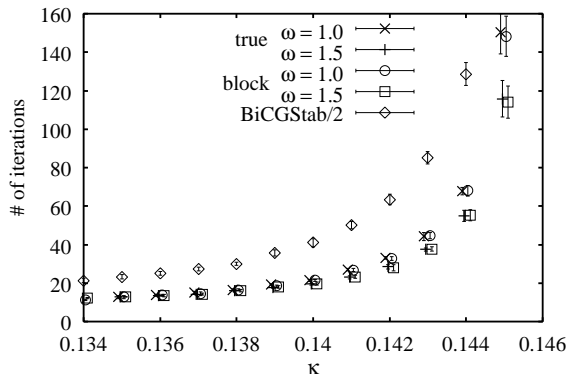}
  \vspace{-10mm}
  \caption{Number of iterations at $c_{SW}=1.0$.
    }
  \label{fig:Clover_1.0}
\end{figure}

Fig.\ \ref{fig:Clover_1.0} shows iteration numbers at $c_{SW}=1.0$ for
both implementations.  BiCGStab/2 represents the estimate for the
odd-even inverter, $\omega = 1.0$ stands for the case without
over-relaxation, the optimal value is $\omega =1.5$. The gain is about
a factor of $2$ against BiCGStab/2 at $\omega =1$; over-relaxation
yields another $10$ to $20\%$ gain. The difference between the two
choices of $D$ is not significant.
\begin{figure}[htb]
  \includegraphics[angle=270,width=\columnwidth]{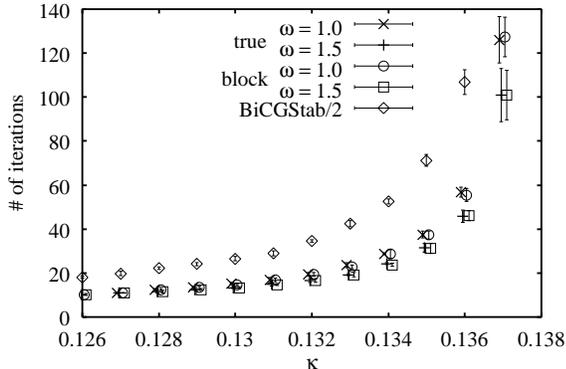}
  \vspace{-12mm}
  \caption{Number of iterations at $c_{SW}=1.6$. 
    }
  \label{fig:Clover_1.6}
  \vspace{-5mm}
\end{figure}
The results shown in Fig.\ \ref{fig:Clover_1.6} for $c_{SW}=1.6$ are
qualitatively identical to the $c_{SW}=1.0$ case.

\section{PRECONDITIONING TPA}
Next we consider a perfect free lattice fermion action for arbitrary
mass \cite{QuaGlu}. As the couplings decay exponentially, a practical
truncation scheme confines the couplings to a unit hypercube
\cite{PA96}.

The matrix for this ``hypercube fermion'' (HF), albeit with $A \propto
{\mathbf 1}$, seems considerably more complicated than the Wilson
fermion matrix, due to contributions of type $C$ and beyond.  But {\it
  ll}\/-SSOR preconditioning and the Eisenstat-trick remain
applicable. We will present a detailed treatement elsewhere
\cite{prep}.

At this stage we discuss the effect of preconditioning by recourse to
the multi-color approach, the extension of the ``red-black'' scheme.
This leads us to $2^{d}$ non-interacting sub-lattices.  We obtain many
off-diagonal blocks that are fortunately largely suppressed.  Denoting
the maximal magnitude of the elements in $L$ and $U$ as $O(\varepsilon
)$, we apply the analog to the odd-even transformation and get
\begin{displaymath}
M' =  \mathbf{1} - ( \sum_{i \geq 1}L^{i}) (\sum_{j\geq 1}U^{j})
= \mathbf{1} - LU - O(\varepsilon^{3}).
\end{displaymath}
We expect the spectrum to be much closer to 1 since the eigenvalues of
$M'$ are all $1-O(\varepsilon^{2})$ (for $M$ they are $1-O(\varepsilon
)$).

\begin{figure}[hbt]
\centerline{\includegraphics[width=.9\columnwidth]{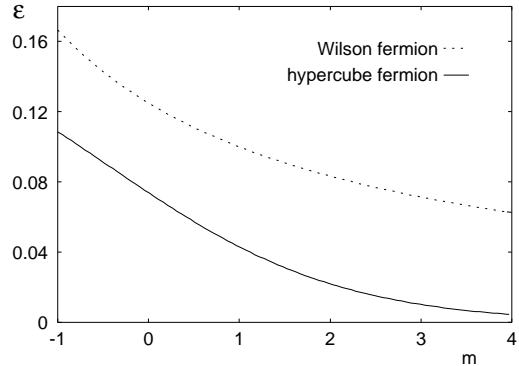}}
  \vspace{-10mm}
  \caption{The off-diagonal magnitude $\varepsilon$ for the
    Wilson fermion ($r=1$) and for the perfect truncated  fermion,
    as a function of the mass.}
  \label{hop}
\end{figure}

Fig. \ref{hop} shows that the parameter $\varepsilon$ obtained is
smaller for the HF than that for the Wilson fermion, since the lattice
derivative is somehow ``smeared over the hypercube''. Thus we expect
multi-color (and also SSOR) preconditioning to work very well.

\section{CONCLUSIONS}

We demonstrated that the application of the {\it ll}\/-SSOR
preconditioning scheme leads to the most efficient preconditioning
known for improved actions.  Our method saves a large factor in memory
compared to odd-even preconditioning.

\end{document}